\documentclass[%
 reprint,
 superscriptaddress,
 amsmath,amssymb,
 aps,
floatfix,
]{revtex4-1}

\usepackage{graphicx}% Include figure files
\usepackage{dcolumn}% Align table columns on decimal point
\usepackage{bm}% bold math
\usepackage{float}
\begin{document}

%\preprint{APS/123-QED}

\title{Role of Structural Morphology in Urban Heat Islands at Night Time}

\author{J.M. Sobstyl}
\affiliation{Concrete Sustainability Hub, Department of Civil and Environmental Engineering, Massachusetts Institute of Technology, Cambridge, Massachusetts, 02139, USA}
\author{T. Emig}
%\email{emig@mit.edu}
\affiliation{%
 MultiScale Materials Science for Energy and Environment, Joint MIT-CNRS Laboratory (UMI 3466), Massachusetts Institute of Technology, Cambridge, Massachusetts 02139, USA 
 }%
 \affiliation{%
 Laboratoire de Physique Theorique et Modeles Statistiques, CNRS UMR 8626, Universite Paris-Saclay, 91405 Orsay cedex, France
 }%
\author{M.J. Abdolhosseini Qomi}
\affiliation{%
 The Henry Samueli School of Engineering, University of California, Irvine, Irvine, CA 92697 
}%
\author{R. J.-M. Pellenq}
\affiliation{Concrete Sustainability Hub, Department of Civil and Environmental Engineering, Massachusetts Institute of Technology, Cambridge, Massachusetts, 02139, USA}
\affiliation{%
 MultiScale Materials Science for Energy and Environment, Joint MIT-CNRS Laboratory (UMI 3466), Massachusetts Institute of Technology, Cambridge, Massachusetts 02139, USA 
 }%
 \affiliation{%
  Centre Interdisciplinaire des Nanosciences de Marseille, CNRS and Aix-Marseille Universite, Campus de Luminy, Marseille, 13288 Cedex 09, France  
}%
\author{F.-J. Ulm}
\affiliation{Concrete Sustainability Hub, Department of Civil and Environmental Engineering, Massachusetts Institute of Technology, Cambridge, Massachusetts, 02139, USA}
\affiliation{%
 MultiScale Materials Science for Energy and Environment, Joint MIT-CNRS Laboratory (UMI 3466), Massachusetts Institute of Technology, Cambridge, Massachusetts 02139, USA
 }%

\date{\today}

\begin{abstract}
  We study the dependence of the intensity of the urban heat island (UHI) on urban geometry. UHI is a urban climate phenomenon referring to the air temperature difference between rural and urban areas.  We use multi-year data for urban-rural temperature differences, combined with building footprint data and a simple heat radiation scaling model to demonstrate for more than 50 cities world-wide that structural morphology -- measured by a building distribution function and the sky view factor -- explains city-to-city variations in nocturnal UHI.  Our results show that the relation between UHI and the morphology is significantly stronger than the one with population, which in the past has been considered as the dominant factor.
\end{abstract}

%\pacs{Valid PACS appear here}

\maketitle

In the century of pullulating global urbanization with 55\% of people living in cities \cite{1}, there is a pressing exigency for establishing quantitative means for controlling climate change \cite{2}. One of the most substantial local climate changes \cite{3}, which has a profound impact on health \cite{4,5} and energy consumption \cite{6} is Urban Heat Island (UHI). While it is well known that the release of solar irradiance heat at night is the inducement of intensified temperatures in cities \cite{7}, the precise role of urban parameters that define the magnitude of UHI remains unknown \cite{9}. Changes in material properties \cite{13}, or morphology of infrastructure \cite{14,15} instigate an alternation of various physical processes at Earth's surface leading to notable climate changes (i.e. UHI). These processes reveal geographical and periodic (i.e. hourly, daily, seasonal) influences on UHI \cite{16,17}. Detailed periodic hourly variations have been found to be related to changes in convection efficiency in the lower atmosphere between different climate zones for the day-time UHI \cite{18}. At night-time, however, it is dominated by two factors: (1) the ability of materials to store solar radiation during the day, and (2) the rate at which this energy is released at night \cite{7}. Additional energy may come in the form of anthropogenic heat \cite{19}, but at night it is reasonable to assume its significance to be negligible.  Energy release of urban morphologies has been related to the sky view from street level, but mostly limited to single street canyons \cite{20}. Contrary to this geometric interpretation, fluctuations of UHI among different cities is considered to be controlled by the population \cite{18}. 

While these factors have been known for several decades, a quantitative understanding of the impact that city texture promulgates on the night-time itensity of UHI, $\Delta T_{u-r}$, remains an open question \cite{9}.  In order to study the impact of structural morphology on UHI, we analyze the hourly night-time peak of $\Delta T_{u-r}$ for twenty-two US urban air temperature time series (in the follwing labelled group A cities) for a period of multiple years \footnote{See Supplemental Material Tab.~I for the group A cities}. The hourly temperature data unveil large fluctuations due to changing weather conditions that superimpose the UHI. However, Fourier transformed temperature series depict distinct maximal peaks at the period of 24 hours (see Fig.~\ref{fig:epsart1}(a), \footnote{See Supplemental Material Fig.~1 for temperature Fourier transforms for group A cities.}). The sum of this peak and the time-averaged temperatures constitutes a reliable measure of the nocturnal $\Delta T_{u-r}$.  

\begin{figure}[tb]
\includegraphics[width=1\linewidth]{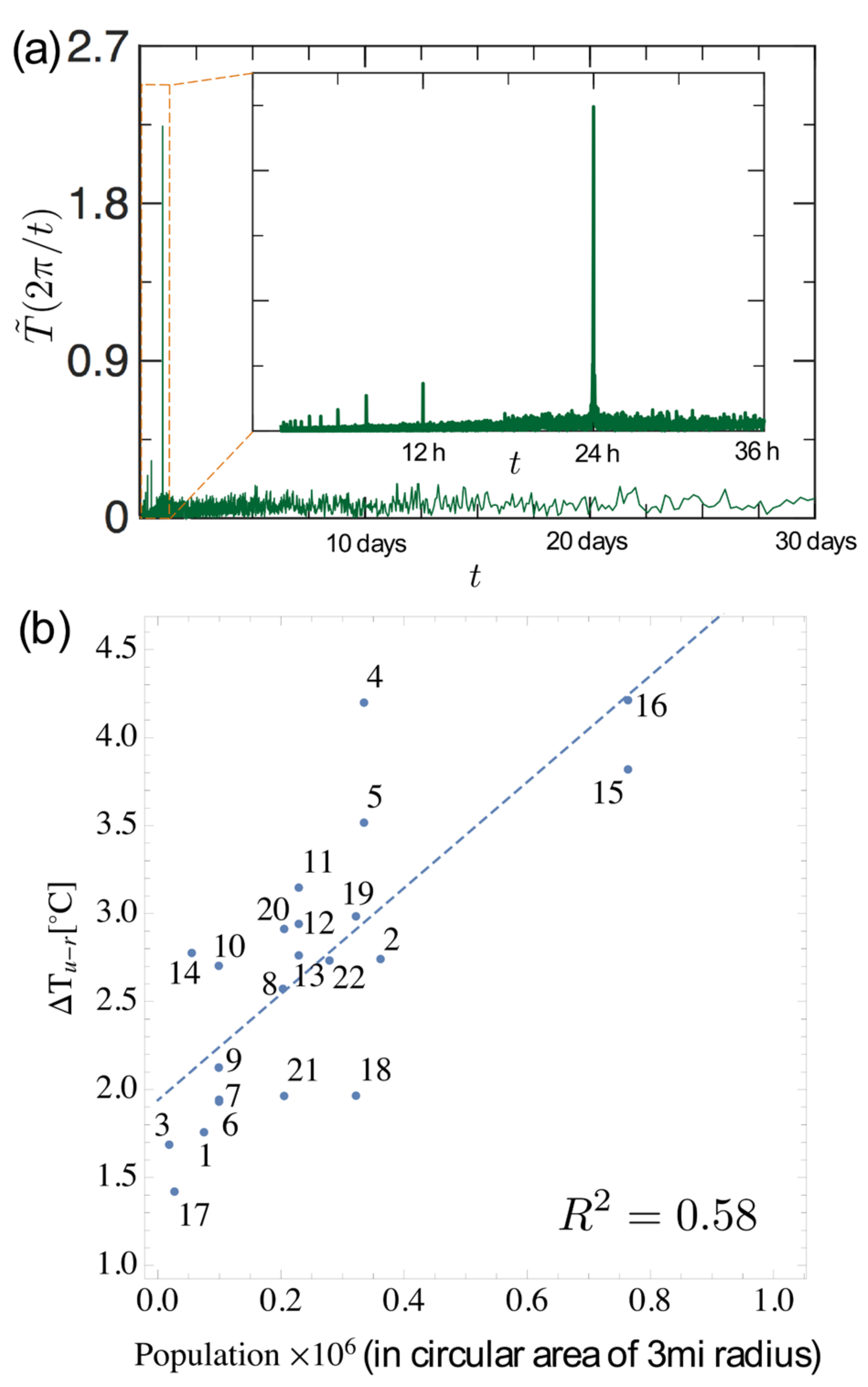} \caption{\label{fig:epsart1} (color online) UHI intensity from Fourier analysis of temperature time series and population influences. (a) Fourier transformed temperature time series (Boston). (b) Relation between $\Delta T_{u-r}s$ and population with 3-mile radius around urban weather station.}  
\end{figure}

\begin{figure}[tb]
\includegraphics[width=\linewidth]{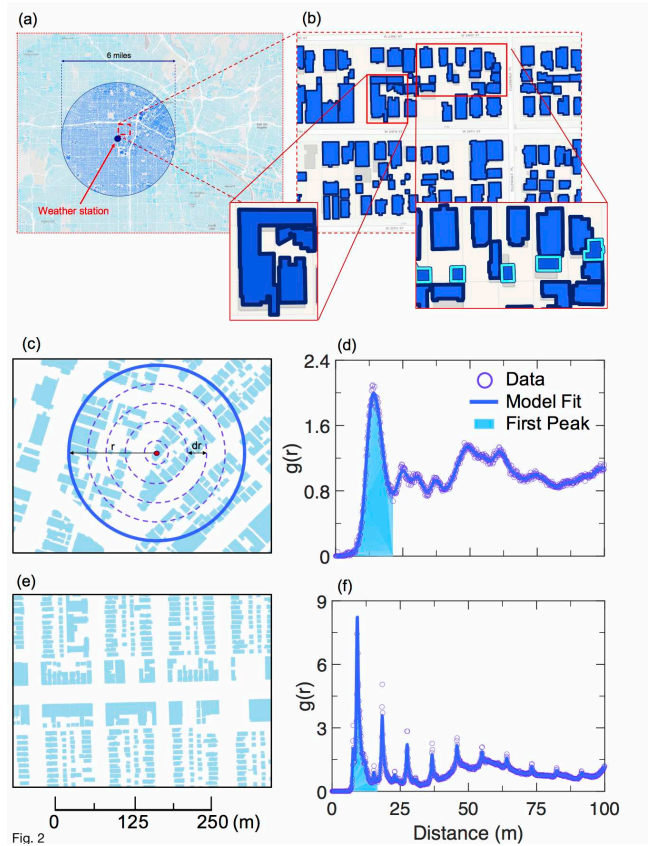}
\caption{\label{fig:epsart2} (color online) Radial distribution function, $g(r)$, for an urban morphology that depicts visualization of data editing, analysis and results.  (a) Buildings within a 3-mile radius of the urban weather station are extracted. (b) Any buildings that share a wall are merged and any unoccupied buildings (i.e. garages) are removed from the sample of buildings transformed into a set of single points. (c) City morphology of Los Angeles, CA showing a comparable absence of order of buildings caused by dispersed streets, which is captured by (d) the smooth and outspread peaks in $g(r)$, which generally are characteristic properties of liquids. (e) City morphology of Chicago, IL showing structure and periodicity, reflected in (f) the sharp and very distinctive peaks of $g(r)$ that are known to be the hallmark of highly ordered and stable crystalline materials. }
\end{figure}

We resort to the device of a radial distribution function, $g(r)$, to extract the prevailing geometrical patterns in cities, defined by the set of building footprints in a three-mile radius around the urban weather station (see Fig.~\ref{fig:epsart2}(a)). Early UHI studies have established empirically a scaling of UHI for thirty-one cities (in the following labelled group B cities, located in North America, Europe and Australia) with population \cite{11} - a common hypothesis for the nocturnal UHI \cite{18}. While our data for $\Delta T_{u-r}$ (see Fig.~\ref{fig:epsart1}(b), \footnote{See Supplemental Material Tab.~II for analysed data.}) are indeed correlated with the population (coefficient of determination $R^2=0.53$, Fig.~\ref{fig:epsart1}(b)), we find that with urban geometry encoded in $g(r)$ there is a much stronger correlation, which matches the robust linear scaling that has been observed between urban geometry for cities of group B and the open sky view factor $\psi_{s}$ ($R^2=0.88$, Fig.~\ref{fig:epsart4}a) \cite{20}.

\begin{figure}[tb]
\includegraphics[width=0.9\linewidth]{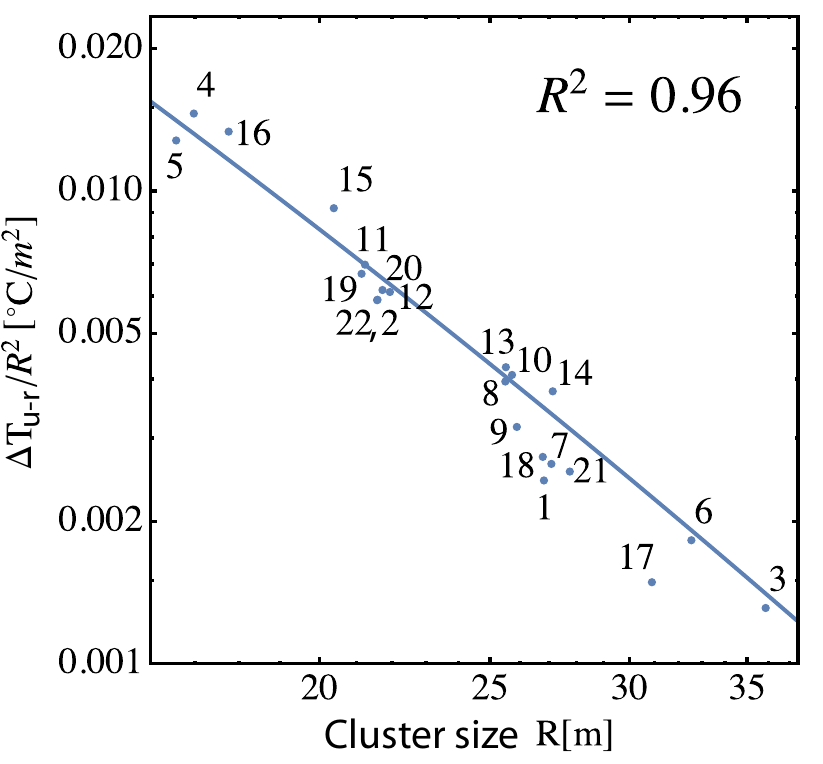}
\caption{\label{fig:epsart3} (color online) Relationship between $\Delta T_{u-r}$ and the cluster size $R$. Measured and model-predicted relationship of urban morphologies obtained from the limits of the integral of the first peak of $g(r)$ shows a strong negative correlation, scaling of which is captured with a power law.}
\end{figure}

Intuitively, a relationship of this kind is consistent with the reduced efficiency at which street canyons release the heat at night \cite{21}. However, our detailed analysis of buildings footprints supports a more complex dependence of UHI itensity on urban morphology. More specifically, utilizing $g(r)$ to quantify urban structure, we find that cities have distinct morphologies resembling structures of crystals, liquids, or other equivalents of other states of matter. For example, we observe that the spatial order varies from liquid-like (Los Angeles, CA, Figs.~\ref{fig:epsart2}c, d) to almost a perfect crystal (Chicago, IL,~Figs. \ref{fig:epsart2}e, f). We use the position of the first minimum of $g(r)$ (Figs.~\ref{fig:epsart2}d, f, \footnote{See Supplemental Material Fig.~3 for $g(r)$ of group A cities.}) to define the local cluster size $R$ and find that its relation to the ratio of  $\Delta T_{u-r}/R^2$ for the cities in group A \footnote{See Supplemental Material Tab.~I for the group A cities.} is consistent with a power law (Fig.~\ref{fig:epsart3}). To reconcile this scaling with the previously established correlation with the sky view, it is instructive to construct a simple heat radiation model. Such a model considers that at night time only long wavelength infrared (IR) radiation emitted from urban surfaces contributes to UHI \cite{22}. To demonstrate that a simple scaling theory accounts for UHI variations with urban structure measured by $g(r)$, we separate contributions of non-geometric origin to $\Delta T_{u-r}$. For that we assume that flat urban surfaces have an average temperature $T_{\rm u,flat}$  that is different from the corresponding temperature of rural surfaces, $T_r$, due to increased sensible heat storage, decreased evapotranspiration and increased absorption of ultra-violet (UV) radiation at day-time \cite{23}. The cumulative effect of the urban-rural difference between the latter processes is summarized by a phenomenological factor $\gamma$  with $T_{\rm u,flat} = \gamma T_r$ for flat surfaces. For a quantitative description of the reduced nocturnal heat release from urban areas due to their increased ``roughness'' we resort to the device of an effective temperature $T_{\rm eff}$ that is often used for a body as an estimate of its surface temperature when the emissivity is unknown \cite{24}. $T_{\rm eff}$ is defined as the temperature of a perfect black body that radiates the same power $P$ as the actual body according to the Stefan-Boltzmann law, $P=\sigma AT^4_{\rm eff}$ ($A$ is the surface area of the body and  $\sigma$ the Stefan-Boltzmann constant, \footnote{See Supplemental Material sections D and E for the description of the heat radiation model.}). By analogy, we apply this concept to cities. Since the wave length of IR radiation is much shorter than all relevant urban length scales, diffraction effects can be neglected and an increase in surface area attributed to buildings (when compared to rural areas) determines $T_{\rm eff}$. Assuming buildings of size $L$, and mean height $\bar{h}$, separated by an average distance $d$ \footnote{See Supplemental Material sections D and E for the description of the heat radiation model.}, our model predicts
\begin{eqnarray}
\label{eq:1}
\Delta T_{u-r}=T_r\Bigg[\gamma \bigg(1+\frac{4L\bar{h}}{(L+d)^2}\bigg)^\frac{1}{4}-1\Bigg] \, .
\end{eqnarray}

\begin{figure}[htb]
\begin{center}
\includegraphics[width=0.9\linewidth]{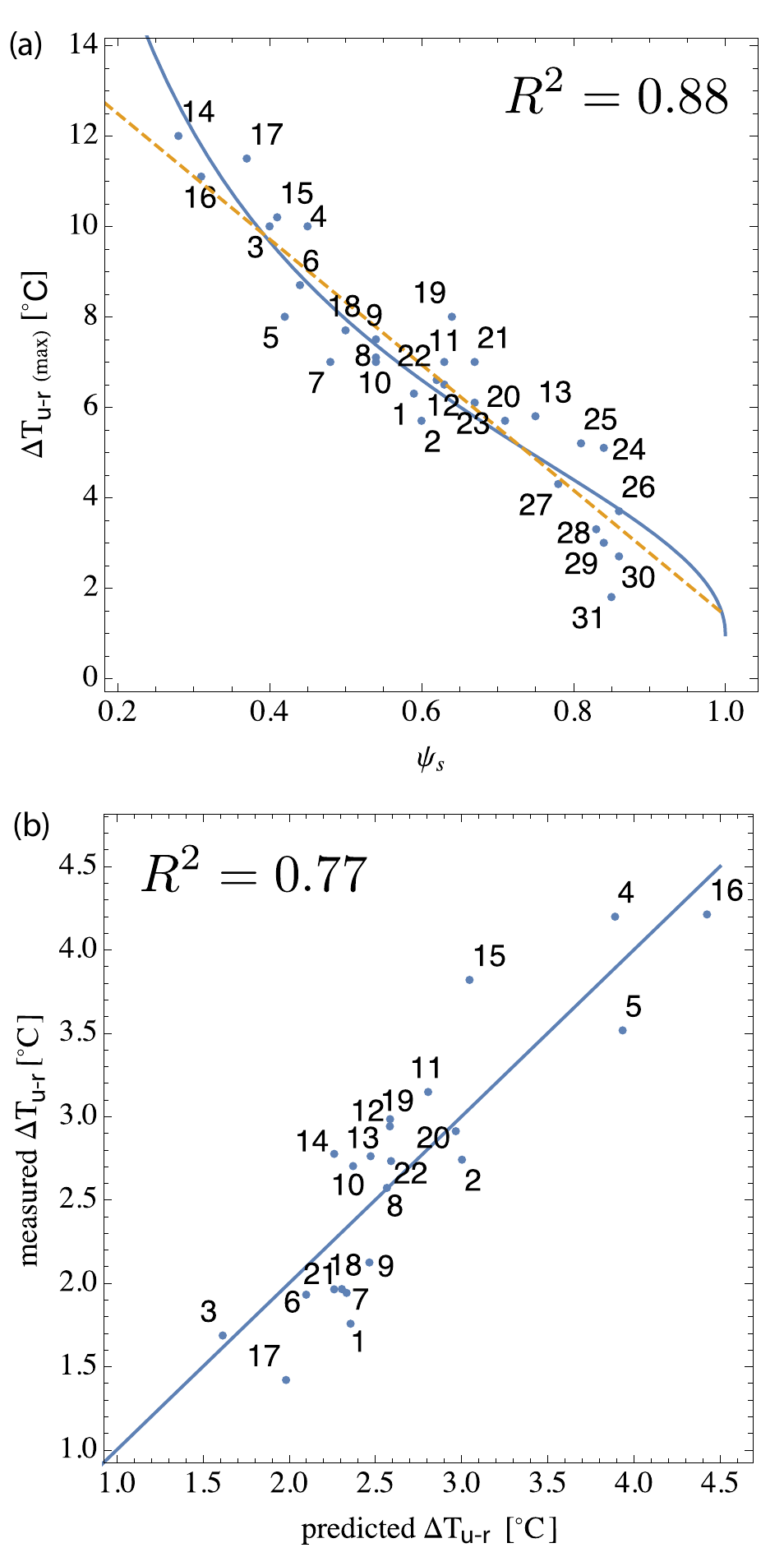}
\caption{\label{fig:epsart4} (color online) Relationship between measured and model-predicted $\Delta T_{u-r}$. (a) maximal $\Delta T$ of city group B as function of the sky view factor $\psi_s$ from Oke \cite{25} together with the linear fit of Oke (dashed line) and the fit to our model (solid curve), see Eq. (2). The numbers refer to the numbering of cities in the work of Oke \cite{25}. (b) Comparison of measured and predicted $\Delta T_{u-r}$ for cities of group A, see Eq. (1).}
\end{center}
\end{figure}

This prediction can be probed by field data in different ways, which is important since the availability of geometric data for most cities are either incomplete (i.e. building heights are missing), or only sky view factors are available. For the data set A, there is no information on building heights. However, detailed information on building footprints is available, thus allowing us to compute $L$ and $d$ \footnote{See Supplemental Material Tab.~II for analysed data.}. We compare these values to our theoretical model by minimizing (with respect to $T_r$, $\gamma$, and $\bar{h}$) the squared deviations between the data for $T_{u-r}$ of all twenty-two cities and the corresponding prediction of Eq.~(\ref{eq:1}) with $L$ and $d$. We find a convincing agreement with the parameters $T_r=20.5^{\circ}$C, $\gamma =1.0$, and $\bar{h}=9.5$m, yielding a coefficient of determination, $R^2=0.77$ (see Fig.~\ref{fig:epsart4}b). Since most of the analyzed urban areas are mainly residential, we conclude that the result for the mean building height $\bar{h}$ is reasonable. However, we have estimated the correction factors for the mean buildings heights that would yield an ideal agreement with our model, showing that corrections of only $\pm30\%$ compared to $\bar{h}$ would be needed for a perfect agreement with Eq.~(\ref{eq:1}).  Knowing the mean height  $\bar{h}$  and building size $L$̄ for all cities of data set A and the mean relation $d = 0.71R$ between the average distance between buildings and the cluster size, Eq.~(\ref{eq:1}) yields a function $\Delta T_{u-r}/R^2$ that can be compared to the measured relation between $\Delta T_{u-r}$ and the cluster size $R$  (see Fig.~\ref{fig:epsart3}). Using $T_r$ as the sole fitting parameter, we find a convincing agreement with $R^{2}=0.96$ for $T_r=24.4^{\circ}C$, which is consistent with the solar radiances values \footnote{See Supplemental Material sections D and E for the description of the heat radiation model.}.

Further credibility of our model is obtained by its application to the previously collected data of group B \cite{25}, providing an insight into $\Delta T_{u-r}$ dependence on building height. We express the ratio $\bar{h}/d$ in terms of the sky view, $\psi_s$, assuming a canyon geometry \cite{20} so that $\bar{h}/d = \frac{1}{2}\tan\big[\arccos(\psi_s)\big]$. Our model then predicts
\begin{eqnarray}
\label{eq:2}
\Delta T_{u-r}(\psi_s)=
T_r\Bigg[\gamma \bigg\{1+\frac{2\frac{L}{d}\tan[\arccos(\psi_s)]}{(1+\frac{L}{d})^2}\bigg\}^\frac{1}{4}-1\Bigg] \nonumber \\ \nonumber \\
\end{eqnarray}
where $T_r$, $\gamma$, and $L/d$ are determined from field data. Contrary to the empirical linear relation between $\Delta T_{u-r}$ and $\psi_s$, Eq.~(\ref{eq:2}) provides an expression that is derived from the fundamental principles of heat radiation. Comparison to the data yields the fitting parameters $T_r=40.4^{\circ}C$, $\gamma = 1.024$, and $L/d = 1.0$ (see Fig.~\ref{fig:epsart4}b), yielding a coefficient of determination of $R^2=0.88$, similar to what has been observed for a linear relation \cite{25}. 
Our analysis thus suggests that city texture plays a key role in determining a city's response to heat radiation phenomena, and points to urban design parameters that can be modulated to mitigate UHI in both the planning and retrofitting of cities \cite{26,27}. In a broader context, our work suggests that tools of statistical physics provide a means to quantitatively address the response of cities to climate. Our results complement previous studies of the factors that influence the day time UHI intensity \cite{18}. The observation that the causes for day and night time UHI are fundamentally different corroborates that $\Delta T_{u-r}$ at day and night time is uncorrelated \cite{28}. According to our findings, the increase of radiating surface area of cities is the main contributor to the nocturnal UHI. While large scale changes to already existing urban morphologies appear unrealistic, efforts of UHI mitigation in the development of future urban structures should aim at minimizing the enveloping surface of urban structures. The resulting reduction in the release of stored heat during night time is expected to have a positive impact on energy consumption and health \cite{4}.

\begin{acknowledgments}
We are grateful for discussions with H.~Van Damme, B.~Coasne and L.~Bocquet. Part of this research was carried out at the Concrete Sustainability Hub at Massachusetts Institute of Technology (CSHub@MIT) with sponsorship provided by the Portland Cement Association (PCA) and the Ready Mixed Concrete (RMC) Research and Education Foundation. Additional support was provided by the ICoME2 Labex (ANR-11-LABX-0053) and the A*MIDEX projects (ANR-11-IDEX-0001-02) funded by the French program ``Investissements d'Avenir'' managed by ANR.
\end{acknowledgments}

%\bibliographystyle{apsrev4-1}

%\bibliography{uhi_paper}

%

\end{document}